\documentstyle{article}

\begin{document}

\input epsf
\def \etal {{\it et al.\/}}
\title{Weak Gravitational Lensing: Current Status and Future Prospects}

\author{NICK KAISER\\
Anglo-Australian Observatory\\
PO Box 296, Epping, NSW 2121, Australia\\
On Leave from CITA, University of Toronto\\
e-mail: {\tt kaiser@cita.utoronto.ca}}

\maketitle

\begin{abstract}
In this review I will describe progress that has been made in determining
masses
of galaxy clusters using `weak lensing' and how this technique my be applied
in the future to determine the dark matter distribution both on supercluster
scales and on the scale of galaxy haloes.
\end{abstract}

\section{Introduction}

Masses of galaxy clusters have traditionally been obtained either
from measurements of the velocity dispersion of the galaxies or from
the temperature of the X-ray emitting gas, and have given a great deal
of input to cosmology: These results have shown that the total cluster
mass greatly exceeds the mass in galaxies and (for any reasonable Hubble
constant) the mass in X-ray emitting gas, and thus suggest the
existence of non-baryonic dark matter.  They have been used to estimate
the density parameter under the assumption that the mass-to-light
ratio of clusters is equal to that of the universe as a whole, and,
more recently, under the assumption that the baryonic-to-total mass
ratio is representative of the universal value.  They have also been
used to try to estimate the cluster mass function, which
potentially has much power to constrain theories for structure
formation.

While vital for cosmology, these mass estimates are subject to a
number of systematic uncertainties arising from the
assumptions that go into the modelling, and which are rather
hard to quantify.  For the virial analysis,
one must make assumptions about the anisotropy of the orbits,
one typically assumes that light traces the mass; that the cluster
is spherical, and one assumes that the cluster is dynamically relaxed.
Similar assumptions go into the X-ray analysis, yet it is not entirely clear
to what extent
these assumptions are valid, or what effect
departures from sphericity etc.~have on the mass estimates.  Furthermore,
any mass estimates made assuming equilibrium conditions must inevitably break
down at
radii $>1-2 h^{-1}{\rm Mpc}$ outside of which material is falling
into the cluster for the first time.  Comparison of X-ray and virial
mass estimates (e.g.~Lubin and Bahcall 1993) show a general correlation,
which is encouraging, but also show a large scatter on a case by case basis
even for well observed clusters, which presumably reflects the breakdown of
one or more of these assumptions.

In contrast, distortion of faint galaxies by the gravitational lens effect
can yield a direct measurement of the 2-dimensional projected
mass density with no assumptions about the dynamical state of the
cluster whatsoever.  Moreover, this technique is applicable
at all radii (and the signal-to-noise is roughly independent of radius
over a wide range of radii for reasonable models for the mass distribution).
For the most massive clusters, this method gives masses to a precision
of better than 20\%, with somewhat higher precision possible
with deeper observations, and promises to give a very powerful
direct measurement of the mass distribution on supercluster scales; around
bright galaxies; and can also provide powerful constraints on the
redshift distribution of very faint galaxies.

\section{Quantitative Cluster Mass Estimation}

The effect of a foreground mass concentration along the line of sight is to
map the surface brightness of distant background objects according to
\begin{equation}
f_{\rm obs}(\vec r) = f_{\rm true}(\vec r -\vec \nabla \phi)
\end{equation}
where $\phi$ is the 2-D dimensionless surface potential which is related to the
surface density by
\begin{equation}
\nabla^2 \phi = 2 \kappa
\end{equation}
where $\kappa \equiv \Sigma / \Sigma_{\rm crit}$.  If we measure
angular position relative to the centre of some galaxy, then
its surface brightness will be distorted according to
\begin{equation}
f_{\rm obs}(\vec r) = f_{\rm true}(\Psi_{ij} r_j)
\end{equation}
where the distortion or amplification tensor is
\begin{equation}
\Psi_{ij} = \delta_{ij} - \partial^2 \phi / \partial r_i \partial r_j
\end{equation}
which is a symmetric $2 \times 2$ tensor.  The effect of a gravitational
lens is to
introduce a spatially coherent distortion of faint background
galaxies.  At small radii the distortion is strong and results in
prominent `giant arcs'.  At larger radii the distortion is weak,
but may still be detected as a statistical anisotropy of the galaxy
shapes.  There are several steps in the process of obtaining a
quantitative mass estimate for a cluster: First one must obtain
high-quality deep photometric data (preferably with multi-colour
information), one must then apply some object finding algorithm
to identify the galaxies and then construct some statistic which
is sensitive to the polarisation of the shapes of the background
galaxies. The next step is to calibrate the relation between the
observed polarisation and the gravitational shear (this depends
on the seeing which will tend to reduce the
apparent polarisation of small objects).  It is then usually necessary to
correct for anisotropy of the instrumental point-spread-function (PSF) as
this can easily introduce spurious polarisation of the faint
galaxies.  At this point one has a noisy 2-dimensional map of the
shear field (or rather a set of noisy estimates of the shear ---
one for each background galaxy), and the next step is to invert
this shear field to obtain the dimensionless surface density
$\kappa = \Sigma / \Sigma_{\rm crit}$.  The final step is to
estimate the critical surface density $\Sigma_{\rm crit}$ and
thereby obtain the physical 2-dimensional surface density $\Sigma(\vec r)$.

\subsection{Shear Measurement}

The most commonly used analysis software is the FOCAS package
(Jarvis and Tyson, 1981) which identifies connected regions lying above some
isophotal threshold (and attempts to split up overlapping galaxy images).
This package calculates a variety of statistics, including intensity
weighted second central moments $I_{ij} = \int d^2 r\; r_i r_j f(\vec r)$,
from which one can form a `polarisation vector'
$\vec e = \{(I_{11} - I_{22}), 2 I_{12}\} / (I_{11} + I_{22})$ whose
expectation value
one can show is proportional to the shear $\vec \gamma =
\{(\phi_{,11} - \phi_{,22}) / 2, \phi_{,12}\}$.  We have developed
an alternative, though rather similar, system (described in Kaiser, Broadhurst
and Squires, 1995; hereafter KSB) which, rather than using an isophotal
threshold,
smooths an image with a hierarchy of smoothing kernels and then
identifies objects as local peaks of the significance as a function
of position and smoothing radius.  Our polarisation is defined
similarly, but using second central moments calculated with a gaussian
radial weight function with scale length matched
to the size of the galaxy in order to minimise the
noise from photon counting statistics.  Another variant is that of
Bonnet and Mellier (1994) who use the square root of the
determinant of $I_{ij}$ in the denominator
in the definition of the polarisation
instead of the trace $I_{11} + I_{22}$.  The common feature of these methods is
that in the absence of lensing, the expectation value of the polarisation
should vanish by symmetry and a coherent shear will introduce a
shift in the mean proportional to $\gamma$.

A variety of techniques have been applied to calibrate the
relation between $\vec e$ and $\vec \gamma$. The basic idea is to figure out
how the polarisation values for a set of galaxies change
under the influence of an applied shear.  The complication is that
the shear is applied before the smearing of the image by the seeing disk,
but a number of methods have been developed to solve this problem.
One approach (Tyson and
Fisher 1995) is to model the properties of the faint galaxies,
and then to generate synthetic images which are sheared by a known
amount and then degraded to mimic the ground based conditions
and then analysed.
A somewhat more direct approach (KSB) is to take deep
HST images of small patches of sky and process these in a similar manner.
While the amount of deep photometry is rather limited at the moment,
such experiments show that relation between the shear and polarisation
is approximately linear, and determine the constant of proportionality
to about 10\% fractional error.  These experiments also allow one
to optimise various features of the analysis to give robust and low
noise estimates of the shear.  Another approach (Wilson, Cole and
Frenk, 1995) is to deconvolve ground based images, shear them and
then reconvolve with the seeing PSF, and they too find that, while the
correction for seeing can be quite large, it can be estimated quite reliably.
Yet another approach is to shear the post-seeing images --- which is
equivalent to shearing the images first and then smearing with a
PSF which is slightly anisotropic --- and then reconvolve with a small
but highly anisotropic PSF designed to recircularise the stellar images.
This problem has also been considered by Villumsen (1995).

In addition to the effect of gravity, the shapes of galaxies are
distorted by any instrumental PSF anisotropy.  Luckily, stars in our
galaxy provide a convenient control sample with which we can measure
the PSF (and provided the seeing is reasonably good
it is relatively straightforward to separate a subsample of faint
stars). If the PSF were constant, it would
then be fairly straightforward to
reconvolve the image with a small but highly anisotropic PSF designed to
recircularise the stars and this would null out any instrumental effect
on the background galaxies.  In fact, the PSF usually tends to vary
across the field, which makes implementing such a scheme rather
difficult.  An alternative which seems to work very well though is to
calculate, for each object, how its polarisation value would respond to
a given PSF anisotropy, and one can
then use the polarisation of the stars to develop a model for the PSF as
a function of position on the image and then apply an appropriate
linearised correction to the galaxy polarisation values.

\section{Direct Estimate of the Convergence}

The steps outlined above provide one with a noisy estimate of the
shear $(\phi_{,11} - \phi_{,22})/2$, from which one must then somehow
reconstruct $\kappa = (\phi_{,11} + \phi_{,22}) / 2$ by one of the
techniques discussed below.
It would be much simpler if one could measure the convergence directly.
This is possible in principle, but somewhat difficult in practice
using only photometric observations at least.
The convergence will brighten and enlarge the faint galaxies
(Broadhurst, Taylor and Peacock, 1995;  Bartelmann and Narayan, 1995).  If
galaxies
had a standard apparent luminosity or standard apparent size one could then
readily determine $\kappa$ directly.  Unfortunately there is a wide
distribution in the apparent sizes, the slope of the faint counts
is such that the `amplification bias factor' nearly vanishes,  and
experiments we have performed using deep HST data similar to those described
above show that there is very little hope of obtaining useful
cluster mass estimates in this way.

A more promising approach has been suggested by Broadhurst (1995) who
has shown that since the very red galaxies have a rather flat counts
slope --- this reflects the fact that as one goes fainter the galaxies
tend to become bluer --- they have a non-negligible (actually negative)
amplification bias and so a cluster will tend to produce a depression in
the counts of such background galaxies.  The method does indeed seem to
work, but tends to be noiser than the shear method
due to the rather low number density of very red galaxies and also the fact
that
these galaxies are clustered.  However, it does provide a useful check on the
results of the shear analysis, and also removes an ambiguity in shear based
mass reconstructions.

\section{Reconstruction Methods}

\begin{figure}
\centerline{
\epsfxsize=300pt \epsfbox{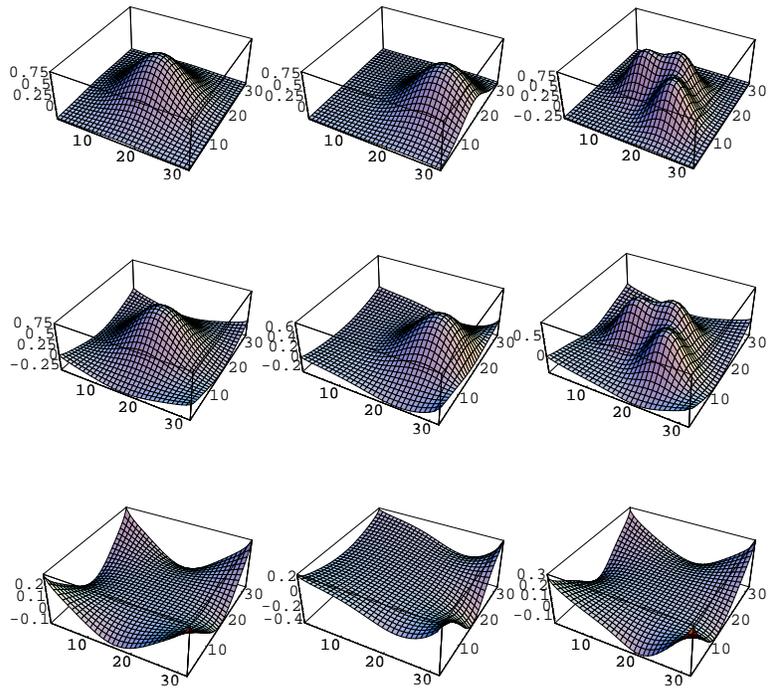}
}
\caption{Illustration of the bias in the Kaiser-Squires '93 method.
The top row of panels show three input model lenses; the
middle panels show the reconstructions, and the bottom row of panels
show the residuals with an expanded vertical scale.}
\label{fig:bias}
\end{figure}

For weak lensing, the desired surface density $\kappa$ and the observed
shear $\vec \gamma$ are all expressed as second derivatives of the surface
potential:
\begin{equation}
\begin{matrix}{
\kappa = (\phi_{,11} + \phi_{,22}) / 2 \cr
\gamma_1 = (\phi_{,11} - \phi_{,22}) / 2 \cr
\gamma_2 = \phi_{,12}
}\end{matrix}
\end{equation}
One can simply fourier transform these equations to obtain algebraic
relations between $\tilde \kappa$, $\tilde \gamma_1$, $\tilde \gamma_2$.
These allow one to recover $\tilde \kappa$ from either  $\tilde \gamma_1$
or $\tilde \gamma_2$ alone (though with particularly bad noise for
certain wave-vectors), and the optimally weighted combination is
\begin{equation}
\tilde \kappa = (\hat k_1^2 - \hat k_2^2) \tilde \gamma_1
+ 2 \hat k_1 \hat k_2 \tilde \gamma_2
\end{equation}
which has a flat noise power spectrum.
On inverse fourier transforming this gives an estimate of $\kappa$ as
a convolution of the shear with a particular kernel:
\begin{equation}
\kappa(\vec r) = \int d^2r'\; K_\alpha(\vec r - \vec r')
\gamma_\alpha(\vec r')
\end{equation}
(Kaiser and Squires 1993).  Simulations (e.g.~Wilson et al., 1995) show that
this works quite well, but there are some limitations.   The kernel
derived above falls asymptotically as $1/r^2$ at large radii, and thus,
strictly speaking, requires data extending to infinity.  With finite
data this results in a bias in the outer parts of the reconstruction
as illustrated in figure \ref{fig:bias}.
Another limitation is that the result is limited to the weak regime.
There has been considerable progress recently in addressing both of these
limitations.

One way to solve the bias problem (Kaiser, 1995) is to exploit the
local expression for the gradient of $\kappa$ in terms of the
gradient of the shear:
\begin{equation}
\vec\nabla\kappa =
\left[\begin{matrix}{
\partial \kappa / \partial x \cr
\partial \kappa / \partial y
}\end{matrix}\right]
=
\left[\begin{matrix}{
\partial \gamma_1 / \partial x +
\partial \gamma_2 / \partial y\cr
\partial \gamma_2 / \partial x -
\partial \gamma_1 / \partial y
}\end{matrix}\right]
\end{equation}
which follows directly from the relations between $\kappa$, $\vec \gamma$ and
$\phi$.  This tells us that in principle at least $\kappa$ is fully
determined up to a constant, and there are many ways one can determine
the surface density at some point relative to its value averaged
over some control region by averaging over line integrals
(Seitz and Schneider, 1995; Schneider, 1995; Kaiser
\etal, 1995).
There are other ways to remove the bias.  It is possible to construct
FFT based methods which are unbiased, but a detailed
comparison of such techniques (Squires and Kaiser, 1995) reveals that
all of the methods so far considered result in somewhat higher
noise in the long-wavelength components (which is unfortunate since
the bias only affects the long wavelengths).

Our currently preferred approach is a regularised
maximum likelihood method (Squires and Kaiser, 1995).  What we do is to
model the surface density as a sum of discrete fourier modes, but on a box
which extends beyond the region where we have data (this is necessary if
one wants the reconstruction to be unbiased).  We can then form the likelihood
$P({\rm data}|{\rm model})$ and one could determine the fourier amplitudes
which maximise this.  If one has many modes (i.e.~good high frequency
resolution) this is unstable, and a certain combination of
modes tends to blow up.  Instead what we do is to apply
a regularisation by minimising $P({\rm data}|{\rm model}, {\rm prior})$
where the prior model is that the mode amplitudes are drawn from gaussian
distributions with variance $P=$ constant.  This very similar to ``classic''
maximum entropy
and is quite simple to calculate.  Now this suffers from the usual
problem with this type of technique that the result tends to be
biased downwards if $P$ is set to be too small.  However, we have
found that in practice one only needs a very mild regularisation (i.e.~large
$P$) to suppress the unstable modes and for a wide range of $P$-values
one obtains a stable and robust reconstruction.  It must be admitted though
that the  results, as with many of the unbiased estimators we have
explored, is rather hard to distinguish from the original biased method.

\subsection{Aperture Densitometry}

The 2-dimensional reconstructions are very nice for some purposes: it is
interesting to compare morphology of the mass and light distribution, and
the fact that the peak of the mass reconstructions coincides with the
centroid of the light gives one some confidence in the method.  However,
if one is interested only in the radial mass profile (or if one simply
wants an estimate of the mass within some radius) then an alternative approach
is to use aperture densitometry.  The basis for this method is a
identity between the mean tangential shear
$\langle \gamma_t \rangle $ taken around a circular
loop and the derivative of the mean surface density within the loop
$\overline \kappa$ wrt loop radius:
\begin{equation}
\langle \gamma_t \rangle = {1 \over 2} {d \overline \kappa \over
d\ln r}
\end{equation}
Integrating this wrt $\ln r$ gives an expression for $\overline \kappa$ within
some aperture of radius $r_1$ relative to the mean of $\kappa$ in some
surrounding annulus $r_1 < r < r_2$:
\begin{equation}
\overline \kappa(<r_1) - \overline \kappa(r_1 < r < r_2)
= \zeta(r_1, r_2) \equiv
{ 2 \over  (1 - r_1^2 / r_2^2)} \int\limits_{r_1}^{r_2} d \ln r
\langle \gamma_t \rangle
\end{equation}
which can then be simply estimated as a sum over the
individual galaxy shear estimates.  A nice feature of this method is that
the result depends only on data outside the aperture, so this can be
chosen to minimise problems with non-linearity and contamination by cluster
members.  Another advantage of this approach (over simply measuring the
mass excess from a 2-D reconstruction say) is that one obtains a simple yet
rigorous estimate of the uncertainty.  This technique was first applied by
Fahlman \etal, (1994), and has also been used by Tyson and Fisher, 1995
in their analysis of A1689.

If one takes $r_2$ to be the outer boundary of the data then one can plot
$\zeta$ vs.~ $r_1$, and thus obtain a lower bound on the mass profile.
One can also compare this with the analogous quantity calculated for
the surface brightness and thereby compare the mass and light profiles
directly.  However, one should bear in mind that that this is then a cumulative
statistic, so the errors at different radii are strongly correlated.

A nice feature of this kind of analysis is that if one applies a
`rotation' $\gamma_1 \rightarrow \gamma_2$, $\gamma_2 \rightarrow -\gamma_1$
to the shear estimates then any real signal should vanish.  This provides
a very useful check on artifacts, or the possibility of spurious
shear arising from intrinsically correlated ellipticities.

\subsection{Non-Linear Reconstruction}

There have been some advances in extending the 2-dimensional mass
reconstruction techniques into the non-linear regime.  If one
only uses the information encoded in the shapes of galaxies, then all one can
hope to learn from observations on some patch of sky is the
ratio and orientation of the eigenvalues of the distortion
tensor $\Psi_{ij}$. Now since $\Psi_{11} = 1 - \kappa - \gamma$,
$\Psi_{22} = 1 - \kappa + \gamma$, it follows that
$\Psi_{22} / \Psi_{11}$ is only a function of the combination
$\gamma / (1 - \kappa)$. Furthermore, we can only observe the
modulus of the ratio of eigenvalues, whereas the actual ratio
flips sign if we cross a critical line.  The upshot of all this
is that one can define an observable polarisation $e = ( 1- R) / (1 + R)$
where $R$ is the ratio of the smaller to the larger (in magnitude)
eigenvalue, which in the general coordinate frame becomes a vector relation,
and which in the even parity regime is equal to $\gamma / (1 - \kappa)$
while in the odd parity regime is equal to $(1 - \kappa)/ \gamma$.
This apparent ambiguity has been dubbed by Schneider and Seitz (1995) a
`local invariance transformation'.

Now using the expression above for the gradient of $\kappa$ in terms
of $\partial \gamma_i / \partial r_j$ it is not difficult to
derive an explicit expression for the gradient of the log of
$1 - \kappa$ in terms of the observable polarisation $\vec e$.
\begin{equation}
\vec \nabla \log (1 -\kappa) \equiv \vec u =
\left[\matrix{
1 + e_1 & e_2 \cr
e_2 & 1 - e_1
}\right]
\left[\matrix{
\partial_x & \partial_y\cr
-\partial_y & \partial_x
}\right]
\left[\matrix{
e_1 \cr
e_2
}\right]
\end{equation}
(Kaiser, 1995).  This is for the even parity region; in the odd parity
region we must use $e = (1 + R) / (1 - R)$.
This expression makes explicit a further ambiguity in non-linear
reconstruction (Schneider and Seitz's `global invariance transformation'):
that one can
always multiply $1 - \kappa$ by some arbitrary constant
without changing the distortion. This is not much of a problem if one
has data extending to large distance, since the `wrong' choice of constant
will give $\kappa$ tending to a non-zero constant at large radii
and will therefore give a mass diverging as $r^2$. It also shows that the
local invariance transformation ambiguity can be resolved: Simply calculate
the curl of the vector $\vec u$; this should of course vanish, but will
only do so in general if one is using the appropriate expression for $\vec e$.

The main practical difficulty we have found in applying this
to real data is contamination of the faint galaxy population by cluster
galaxies in the central parts of the cluster.

\section{Determining the Critical Surface Density}

The above methods allow one to determine $\kappa$, the dimensionless
surface density.  In order to obtain the physical surface
density one must multiply by $\Sigma_{\rm crit}$
which depends on the redshift distribution for the faint galaxies.
Now for a single screen of sources, and assuming a Einstein - de Sitter
universe, the critical surface density is
$\Sigma_{\rm crit} = (4 \pi a_l w_l \max(1 - w_l / w_s, 0))^{-1}$,
where comoving distance $w = 1 - 1/\sqrt{1 + z}$.  In general, if
we average over galaxies with some distribution of redshifts
we must use the  $\Sigma_{\rm crit, effective} =
1 / \langle 1/\Sigma_{\rm crit} \rangle$ in place of
$\Sigma_{\rm crit}$. The big problem is that the redshift distribution
is rather poorly known (the deepest complete surveys
only reach magnitudes $I  \simeq 22$ (Lilly \etal, 1995), whereas the typical
lensing observations go 2 or more magnitudes deeper.
Now for low redshift lenses, $z_l \sim 0.2$, the effective critical
surface density is very weakly dependent on $n(z)$ and the uncertainty
in determining $\Sigma_{\rm crit}$ is  very small
($\sim 10\%$).  However, for more distant clusters (and for some
of the other applications described below) $n(z)$ is a big issue.

In some of our cluster observations we can detect the shear for
galaxies in the magnitude range where redshifts are available,
albeit with less precision than if we use all the galaxies,
and by comparing with the shear strength observed
for the fainter samples we can constrain $n(z)$ at fainter magnitudes.
With only one cluster, this is rather noisy, but if one had a sample of
say 10 or so massive clusters at the same redshift then
one can hope to constrain $n(z)$ for faint galaxies quite well,
and thus tighten up the normalisation of the absolute mass scale.  By studying
cluster lenses at a range of redshifts one can hope to build up
a fairly detailed picture of $n(z;m)$; this, rather than
cluster mass estimation,  has been the primary
goal of the Durham group (Smail \etal, 1995), who targeted
relatively high-$z$ clusters, and  the idea
has been discussed by Bartelmann and Narayan (1995).
An exciting, if somewhat distant,
prospect is that if one could obtain complete samples of spectroscopic
redshifts at these magnitudes then by comparing with the
lens inferred estimates (which really measure the
relative comoving distance distributions), one might be able
to constrain the cosmological world model.

\section{Results for Clusters}

\begin{figure}
\centerline{
\epsfxsize=200pt \epsfbox{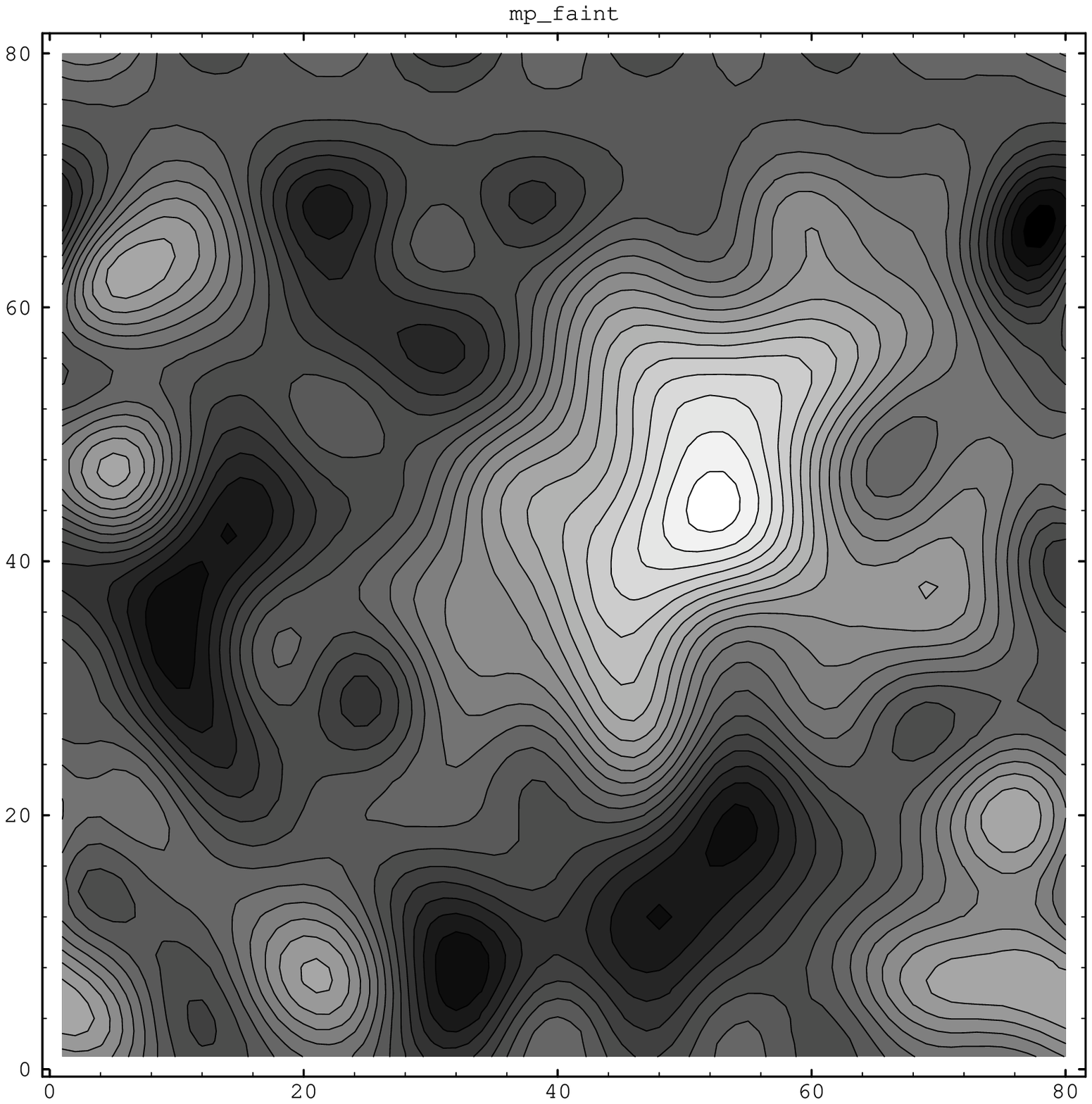}
\epsfxsize=200pt \epsfbox{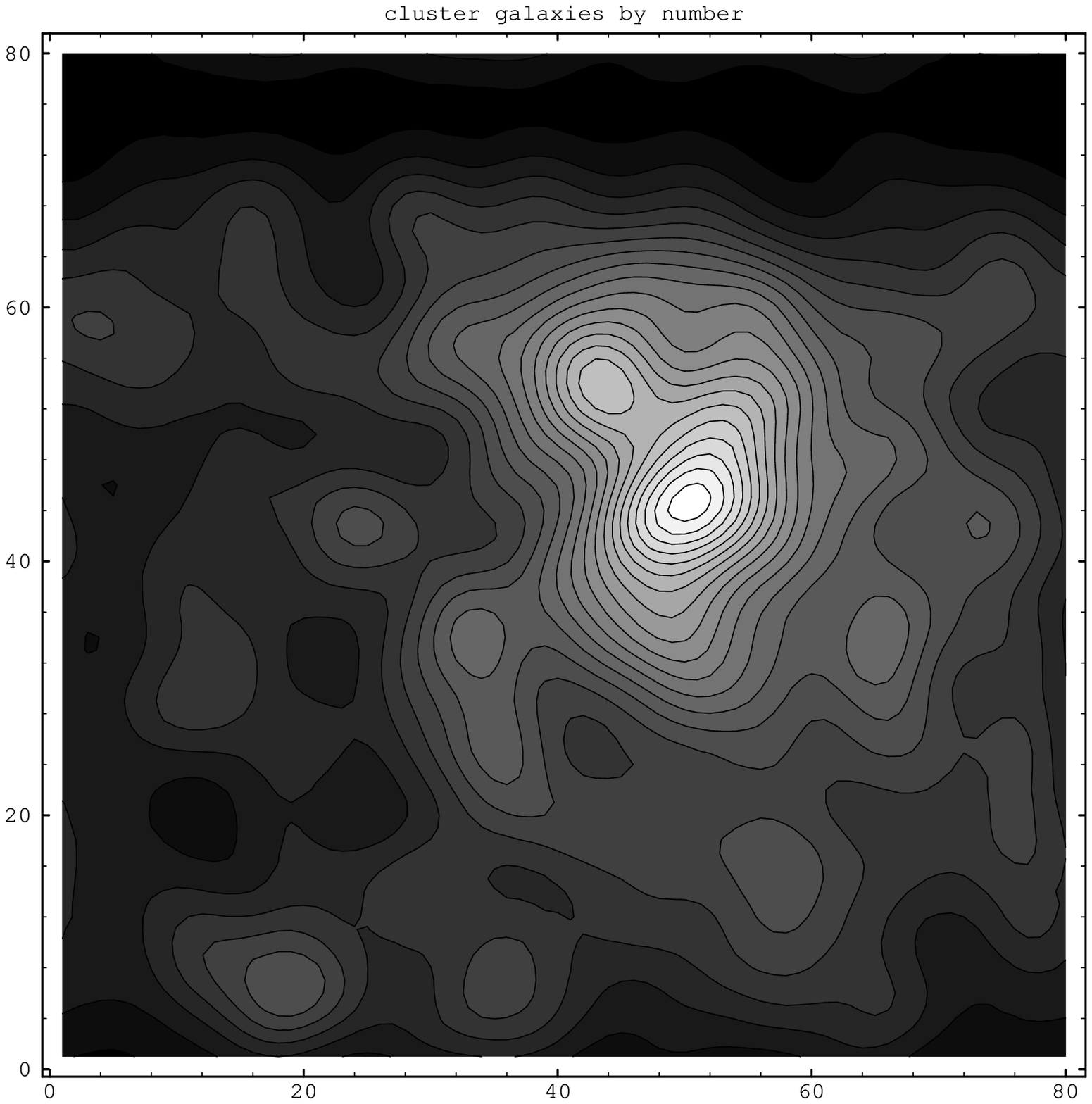}
}
\caption{The left panel shows the reconstruction of the surface
mass density for A1689 made using the `regularised maximum likelihood'
method.  The panel on the right shows the surface number density
of cluster galaxies.  The side of the box is $\sim 8'$ or
about $1 h^{-1}$Mpc at the redshift of the cluster.}
\label{fig:2dreconstruction}
\end{figure}

\begin{figure}
\centerline{
\epsfxsize=200pt \epsfbox{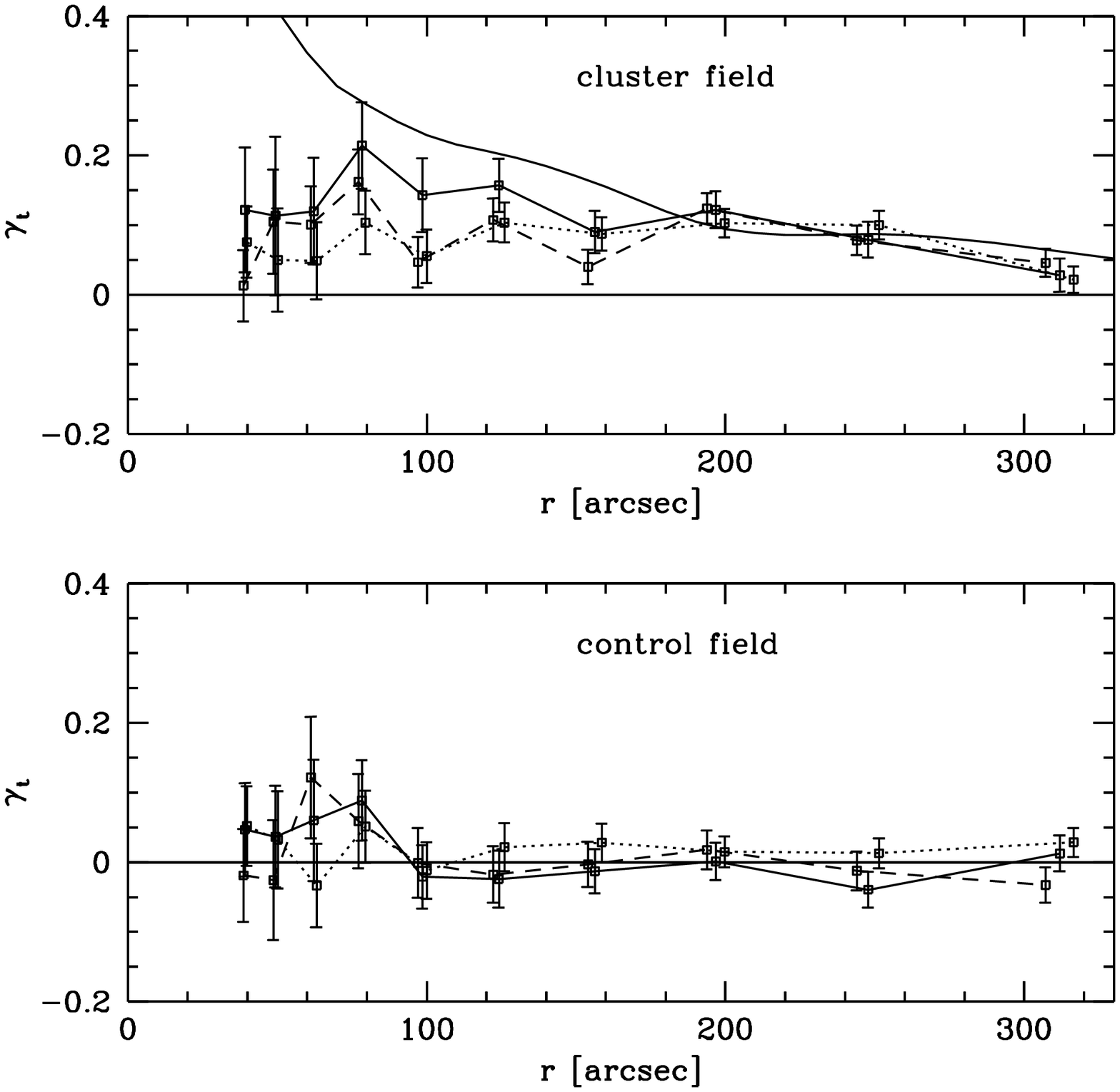}
\epsfxsize=200pt \epsfbox{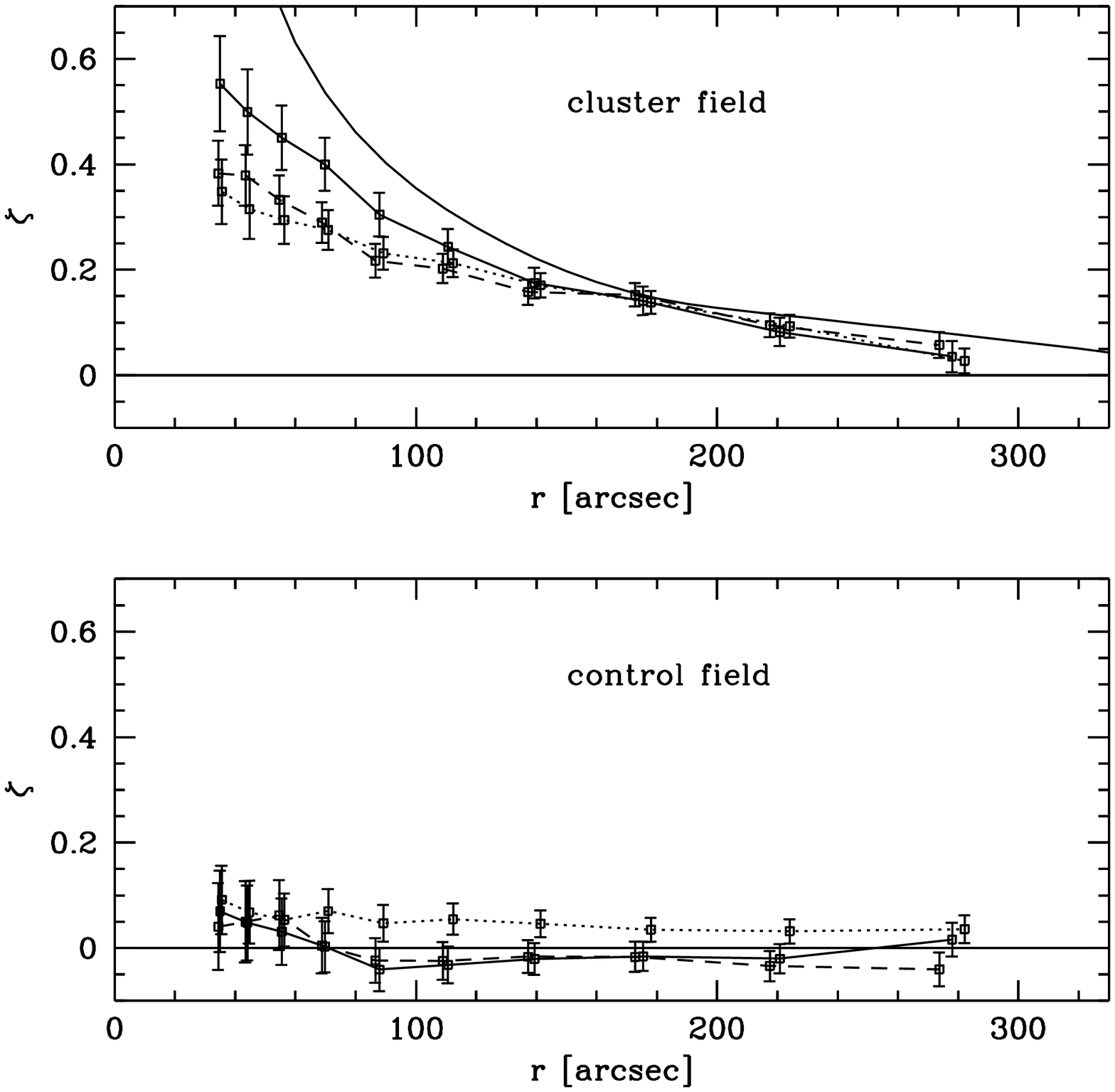}
}
\caption{A1689 tangential shear profile is shown on the left.
Upper plot shows the tangential
shear profile with V-band dotted, I-band dashed and the color selected
sample solid. The smooth solid line is the predicted shear
for a mass-to-light ratio of $450 h$ as inferred from the giant
arc. Lower plot shows same
for control field. The $\zeta$ statistic used for calculating mass within
an aperture is shown on the right (beware that the errorbars in $\zeta$
are correlated).}
\label{fig:etplot}
\end{figure}

The 2-dimensional surface density reconstruction for
A1689 is shown in figure \ref{fig:2dreconstruction} along with
the projected density of cluster galaxies.   The peaks of the
mass and light coincide very well indeed
and there seems to be some similarity in the morphology of the
mass and light distributions. The data used here were taken by Tom Broadhurst
at the
NTT and comprise about 3 hrs net integration in both V and I
passbands.  He also observed a control field which gives a
useful check on the reliability of the analysis.

A more quantitative picture of the shear
profile is shown in figure \ref{fig:etplot}.  These figures show
that the shear is measurable repeatably in both passbands, and
that the strength of the shear at large radii agrees with
that predicted for a mass-to-light ratio of about 450 $h$ in solar units.
(Tyson and Fisher found essentially the same value, though
their calibration involved modelling of the faint galaxy
population rather than degraded HST images as used here).
It is encouraging that no significant shear is found in the
control field. Closer to the cluster centre, the measured shear is
much smaller than that predicted.  This is largely due to
contamination by cluster members.  The solid line with errorbars
shows the result of removing galaxies which lie in the conspicuous
horizontal sequence on the colour-magnitude diagram.   This increases
the shear substantially; the residual discrepancy may be due
to the mass having a somewhat flatter profile than the cluster galaxies,
but it may also be due to bluer cluster members we have failed to remove.

We have performed a similar analysis on A2218 (Squires, \etal, 1995),
and obtain a similar mass-to-light ratio, though with
somewhat larger ($\sim 20\%$) statistical uncertainty (for A1689
the statistical error in the aperture mass is only about 10\%,
with a systematic error of about the same order arising
from uncertainties in $n(z)$ etc.).  The first cluster
we applied this technique to, ms1224, gave a somewhat
larger $M/L \simeq 800 h$ (and a mass much larger than obtained
by virial analysis).  From early, but rather limited, data on A2163,
the hottest X-ray cluster known, we found surprisingly
little evidence for shear. We now have much more data on this
cluster, but have yet to analyse it in detail.

Smail \etal 1995, have obtained broadly similar results for
several clusters, obtaining $M/L \simeq 500 h$, and the
Toulouse group (Bonnet \etal, 1993; Bonnet \etal, 1994)
also measured the weak shear field around Q2345+007 and in the
cluster CL0024.

The mass-to-light ratios from weak lensing are somewhat
larger than those typically obtained from X-ray or virial
analysis (though as mentioned, in some cases the lensing mass
falls below the other estimates).  It is then natural to try to
estimate $\Omega$ (under the usual, though obviously
questionable assumption that the $M/L$ for the cluster is
that same as that for the universe as a whole).  If one uses
the estimates for the comoving luminosity density measured
locally (e.g.~Loveday \etal, 1992) then one infers $\Omega \simeq 0.3$.
However, there are strong indications from recent fainter
redshift surveys (Lilly \etal, 1995;   Colless, 1995)
that the comoving luminosity density is actually much higher, and then
this exercise gives $\Omega \simeq 1$.

An interesting new approach is to target bright radio sources rather than
cluster
lenses (Fort \etal, 1995).  The idea is that these very bright objects may have
been
amplified by intervening matter.  If so, and if the lensing structures
lie at $z < 0.5$ or so, then one might be able to detect the shear
due to the hypothetical lens on faint galaxies, and the data do indeed show
the expected signal.

\section{Future Prospects}

This subject is currently undergoing a major boost in instrumentation.
In the past the size of detectors has been very limited, and in order
to measure the shear at large radii from the cluster centre has required
painstakingly building up a mosaic of images.  CFH has recently
introduced MOCAM (a mosaic of 4 chips each the same size as the earlier
$2048^2$ FOCAM chip), and this will shortly be superseded by
Luppino's mosaic of eight $2048 \times 4096$ chips which is
4 times larger still (Metzger, Luppino and Miyazaki, 1995).
With these new instruments it will be possible to
measure the shear in clusters to much larger radii and/or
push deeper for better signal to noise.  In addition several
new possibilities open up.

One possibility is to use galaxy-galaxy lensing to constrain the
profiles of dark haloes around galaxies.  Individual bright galaxies
with rotation speeds of $\simeq 200$ km/s
are $\sim 100$ times weaker lenses than very massive clusters like
A1689.  As the latter was only detected at about the 10-sigma level,
bright galaxies are clearly too weak to detect individually.  However,
by stacking the results for many foreground galaxies it may
be possible to build up a statistical halo profile.  This was
originally tried by Tyson \etal (1984) using photographic plates.  We have
looked for the effect with our limited number of blank control fields etc.~but,
not too surprisingly, find a null detect as yet.  There have been
two recent claims to detect the effect: Brainerd \etal (1995), using a single
10'-square
field taken at Palomar, and the MDS team have found the effect in HST
images of the `Groth Strip'.
Both of these results are of fairly low statistical significance, but with
much more extensive photometry that should now become feasible the
situation should improve dramatically.

Another promising approach is to determine the power-spectrum of
mass-fluctuations
on supercluster scales.  The first quantitative predictions of weak lensing
by large-scale structure were made by Blandford \etal, 1991, and
Miralda-Escude,
1991.  The predictions are somewhat model dependent, but a
shear at about the 1\% level and coherent over degree scales seems inescapable.
A nice way to analyse this kind of observations is to measure the
2-D power spectrum of the shear, which is then related in a rather
simple and direct manner to the 3-D power spectrum of mass fluctuations
$P(k)$ (Kaiser, 1992).
Such a signal is well within reach: our individual cluster fields each give
null results for the net shear at about the 1\% level, but cover only
small fraction of a square degree (see also Mould \etal, 1994;
Villumsen, 1995).  By increasing the field size it should
prove possible to detect the effect at a high level of confidence
on each degree scale `pixel' and
thus put powerful constraints on theories for structure formation.

\section{References}

\noindent
Lubin, L., and Bahcall, N., 1993. Ap.J., 415, L17

\noindent
Jarvis, J.F., and Tyson, J.A., 1981. AJ, 96, 476

\noindent
Kaiser, N., Squires, G., and Broadhurst, T., 1995. ApJ, in press

\noindent
Bonnet, H., and Mellier, Y., 1994. submitted to A\&A

\noindent
Tyson, J.A., and Fisher, P., 1995, ApJ, 446, L55

\noindent
Wilson, G., Cole, S., and Frenk, C., 1995, preprint

\noindent
Villumsen, J.V., 1995. preprint, astro-ph/9507007

\noindent
Villumsen, J.V., 1995. preprint, astro-ph/9503011

\noindent
Broadhurst, T., Taylor, A.,  and Peacock, J., 1995. ApJ, 438, 49

\noindent
Bartelmann, M., and Narayan, R., 1995. ApJ, in press, astro-ph/9411048

\noindent
Broadhurst, T., 1995. in proceedings ``5th Maryland Dark Matter Meeting'',
astro-ph/9505010

\noindent
Kaiser, N., and Squires, G., 1993. ApJ, 404, 441

\noindent
Kaiser, N., 1995. ApJ, 439, L1

\noindent
Seitz, S., and Schneider, P., 1995. A\&A, in press, astro-ph/9503096

\noindent
Schneider, P., 1995. A\&A, in press, astro-ph/9409063

\noindent
Kaiser, N., Squires, G., Fahlman, G., Woods, D. and Broadhurst, T., 1995.
In ``Wide Field Spectroscopy and the Distant Universe'',
proc. 35th Herstmonceux Conference, ed. S. Maddox, World Scientific

\noindent
Squires, G.,  and Kaiser, N., 1995, in preparation

\noindent
Fahlman, G., Kaiser, N., Squires, G., and Woods, D., 1994. ApJ, 437, 56

\noindent
Schneider, P. and Seitz, C., 1995, A\&A, 294, 411

\noindent
Smail, I., Ellis, R.S., Fitchett, M.J.\ \& Edge, A.C., 1995. MNRAS, 273, 277

\noindent
Squires, G., Kaiser, N., Babul, A., Fahlman, G., Woods, D.,
Neumann, D., \& Bohringer, H. 1995.  To appear in the ApJ.

\noindent
Bonnet, H., Fort, B., Kneib, J.-P., Mellier, Y., Soucail, G., 1993. A\&A, 280,
L7

\noindent
Bonnet, H., Mellier, Y., Fort, B., 1994. ApJ, 427,  L83

\noindent
Loveday, J., Peterson, B., Efstathiou, G., and Maddox, S.J., 1992. ApJ, 390,
338

\noindent
Lilly, S., Tresse, L., Hammer, F., Crampton, D., and Le Fevre, O., 1995.
preprint, astro-ph/9507079

\noindent
Colless, M., 1995. In ``Wide Field Spectroscopy and the Distant Universe'',
proc. 35th Herstmonceux Conference, ed. S. Maddox, World Scientific

\noindent
Fort, B., Mellier, Y., Dantel-Fort, M.,Bonnet, H., Kneib, J.-P.: 1995,
preprint,
astro-ph/9507076

\noindent
Metzger, M., Luppino, G., and Miyazaki, S., 1995. preprint

\noindent
Tyson, A., Valdes, F., Jarvis, J., and Mills, A., 1984. ApJ, 281, L59.

\noindent
Brainerd, T., Blandford, R., and Smail, I., 1995. preprint, astro-ph/9504010

\noindent
Blandford, R., Saust, A., Brainerd, T., and Villumsen, J., 1991. MNRAS, 251,
600

\noindent
Miralda-Escude, J., 1991. ApJ, 380, 1

\noindent
Kaiser, N., 1992. ApJ, 388, 272

\noindent
Mould, J.,, Blandford, R., Villumsen, J., Brainerd, T., Smail, I. et al.:
1994,
ApJ, 271,  31

\end{document}